\documentclass{sigchi}

\CopyrightYear{2017} 
\setcopyright{acmlicensed} 
\conferenceinfo{UIST '17 Adjunct,}{October 22--25, 2017, Quebec City, QC, Canada}
\isbn{978-1-4503-5419-6/17/10}
\acmPrice{\$15.00}
\doi{https://doi.org/10.1145/3131785.3131842}

\usepackage{balance}       
\usepackage{graphics}      
\usepackage[T1]{fontenc}   
\usepackage{txfonts}
\usepackage{mathptmx}
\usepackage[pdflang={en-US},pdftex]{hyperref}
\usepackage{color}
\usepackage{booktabs}
\usepackage{textcomp}

\usepackage{tikz}

\usepackage{microtype}        
\usepackage{ccicons}          

\usepackage{todonotes}

\teaser{
\centering
\vspace{-3mm}
  \includegraphics[width=1.0\linewidth]{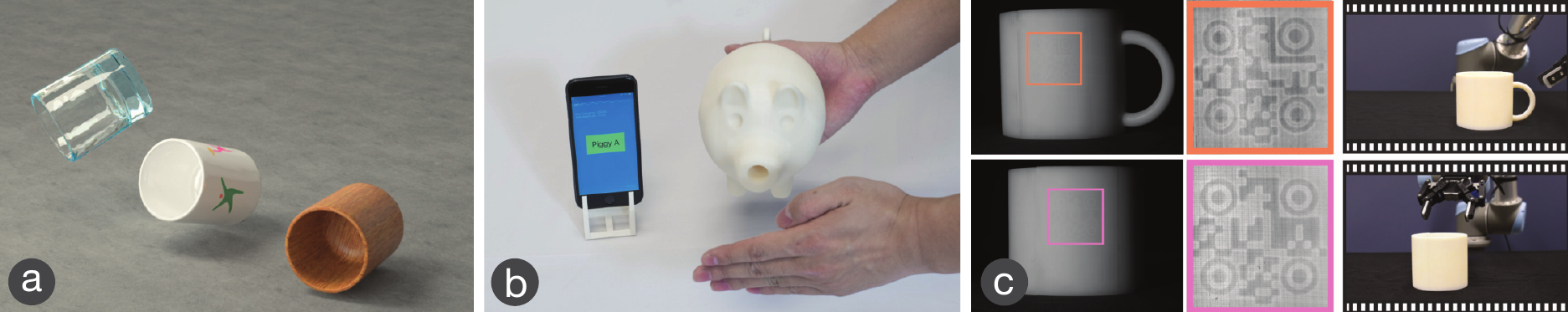}
  \caption{
  (a) Interactively physics-based sound simulator helps artists try out different 
      materials and auralize the results immediately.
  (b) An example of Acoustic Voxels. By optimizing how the internal cavity filters  sound waves,
  we can embed unique acoustic signatures in different models.
  (c) AirCode invisible tags are designed with our physics-based light scattering tool.
  Here the tags not only  identify the mug but also help estimate its pose, leading to
  a successful robotic grasp.
   \label{fig:teaser}
   }
}

\def\plaintitle{{Interacting with Acoustic Simulation and Fabrication}}

\def\emptyauthor{}
\def\plainkeywords{3D printing; computational fabrication; virtual reality; interaction; design tools; physics-based simulation; audio;}

\makeatletter
\def\url@leostyle{%
  \@ifundefined{selectfont}{
    \def\UrlFont{\sf}
  }{
    \def\UrlFont{\small\bf\ttfamily}
  }}
\makeatother
\def\pprw{8.5in}
\def\pprh{11in}

\definecolor{linkColor}{RGB}{6,125,233}
\hypersetup{%
  pdftitle={\plaintitle},
  pdfauthor={\emptyauthor},
  pdfkeywords={\plainkeywords},
  pdfdisplaydoctitle=true, 
  bookmarksnumbered,
  pdfstartview={FitH},
  colorlinks,
  citecolor=black,
  filecolor=black,
  linkcolor=black,
  urlcolor=linkColor,
  breaklinks=true,
  hypertexnames=false
}

\usepackage{xstring}
\def\FormatName#1{%
  \IfSubStr{#1}{Dingzeyu}{{{\bf #1}}}{#1}%
}

\begin{document}

\title{\plaintitle}

\numberofauthors{1}
\author{%
  \alignauthor{Dingzeyu Li\\
    \affaddr{Columbia University}\\
    \email{dli@cs.columbia.edu}}\\
}

\maketitle

\begin{abstract}
  Incorporating accurate physics-based simulation into interactive design tools 
  is challenging.
  However, adding the physics accurately becomes crucial to several emerging technologies.
  For example, in virtual/augmented reality (VR/AR) videos,
  the faithful reproduction of surrounding audios is required to bring the 
  immersion to the next level.
  Similarly, as personal fabrication is made possible with accessible 3D printers,
  more intuitive tools that respect the physical constraints can help artists
  to prototype designs.
  One main hurdle is the sheer amount of computation complexity to accurately
  reproduce the real-world phenomena through physics-based simulation.
  In my thesis research, I develop interactive tools that implement efficient 
  physics-based simulation algorithms for automatic optimization
  and intuitive user interaction. 
  

\end{abstract}

\category{H.5.m.}{Information Interfaces and Presentation
  (e.g. HCI)}{Miscellaneous} 


\section{Introduction}

With the increasing accessibility of mixed reality devices and consumer-level 
3D printers, recent technology enables us to experience a more 
immersive virtual world as well as fabricate personalized digital models.
The immersion in virtual reality (VR) and the interactions with personalized 3D models
require new design tools that respect the underlying physics to 
keep the seamless immersion.
For example, it is important to provide accurate audio propagation in VR scenes
in order to achieve full immersion.
My thesis research focuses on physics-based design tools that help
users achieve desired functionalities. 

During the design process, interactive feedback is crucial, 
since it not only gives a quick updated view for the edits 
but also guides the trial-and-error improvement process. 
For non-intuitive and sometimes complex physical phenomena, 
for example, sound propagation or resonant chamber design, 
one would usually resort to accurate and predictive simulations. 
However, it is challenging to achieve interactive performance while  
obtaining accurate simulation results. 
In my research, I develop tools based on physical principles,
augmenting design tools with simulations.

My first project is an interactive tool that allows the user to explore different
materials for animations and outputs synchronized sounds accurately 
(Figure~\ref{fig:teaser}-a).  
I designed algorithms to accelerate the computation for sound propagation and 
implemented an efficient runtime approximation scheme, 
achieving realistic audio in virtual scenes. 
With previous methods, if users want to listen
to an updated sound with slightly different material parameters (e.g. Young's modulus),
the computation would take minutes or even longer. 
We validated our interactive tool through numerical experiments and user studies~\cite{Li:2015:transfer}.

Another area where physics-based tools are helpful is computational design for fabrication.
Manipulating 3D geometry with desired properties is difficult since the relationship
between geometry and physical properties is very non-intuitive.
Acoustic Voxels is a system that predicts and optimizes the internal structure to
meet the resonant frequency requirements~\cite{Li:2016:acoustic_voxels}. 
This tool enables users to explore acoustic filters with different shapes, 
creating musical instruments in unconventional 
shapes and motivating new applications in data encoding (Figure~\ref{fig:hippo},\ref{fig:oct}).
Take the encoding idea one step further, I propose AirCode, an
unobtrusive tagging tool to design and embed small air pockets beneath the 
surface~\cite{aircode}.
These AirCode tags can be 3D printed easily without extra processing and detected 
using our consumer-level camera system.

By exploring various tools for different design tasks, 
my thesis aims to bring physics-based simulation into design tools.
In the following, I review related literature, 
present my research projects on physics-based tools, 
and discuss future research directions.


\section{Related Work}

Two themes of existing research are mostly related to my proposed area:
(i)  interactive design tools; and 
(ii) computational design methods for personalized fabrication.
Here I only discuss several representative projects
among abundant literature under these two themes.

\paragraph{Interactive Design Tools}

To ease the design process, various interactive 
tools have been proposed.
For example, in order to add sensors into 3D printed objects,
Capricate designs custom-shaped sensors that fit on complex surfaces and automatically
wire the underlying sensors~\cite{SchmitzKBLMS15}.
Another design tool, aeroMorph, focuses on simulating bending mechanism that creates
shape-changing behaviors with common materials, such as paper and plastics~\cite{OuSVHCPI16}.
With the interactive visualization, users receive feedbacks interactively as they work
on the design. This tool provides great flexibility for them, bypassing 
time-consuming traditional validation approaches which require repeated fabrication.
Platener, a low-fidelity fabrication tool, shows users an interactive interface with a
global slider to define the fidelity-speed trade-off, making it easier to
decide on fabrication fidelity~\cite{BeyerGMCB15}.
My goal is to develop interactive tools and augment them with efficient physics-based
simulations and optimization, providing more accurate interactive feedback on complex 
problems.

\begin{figure}[t]
  \centering
  \includegraphics[width=0.92\columnwidth]{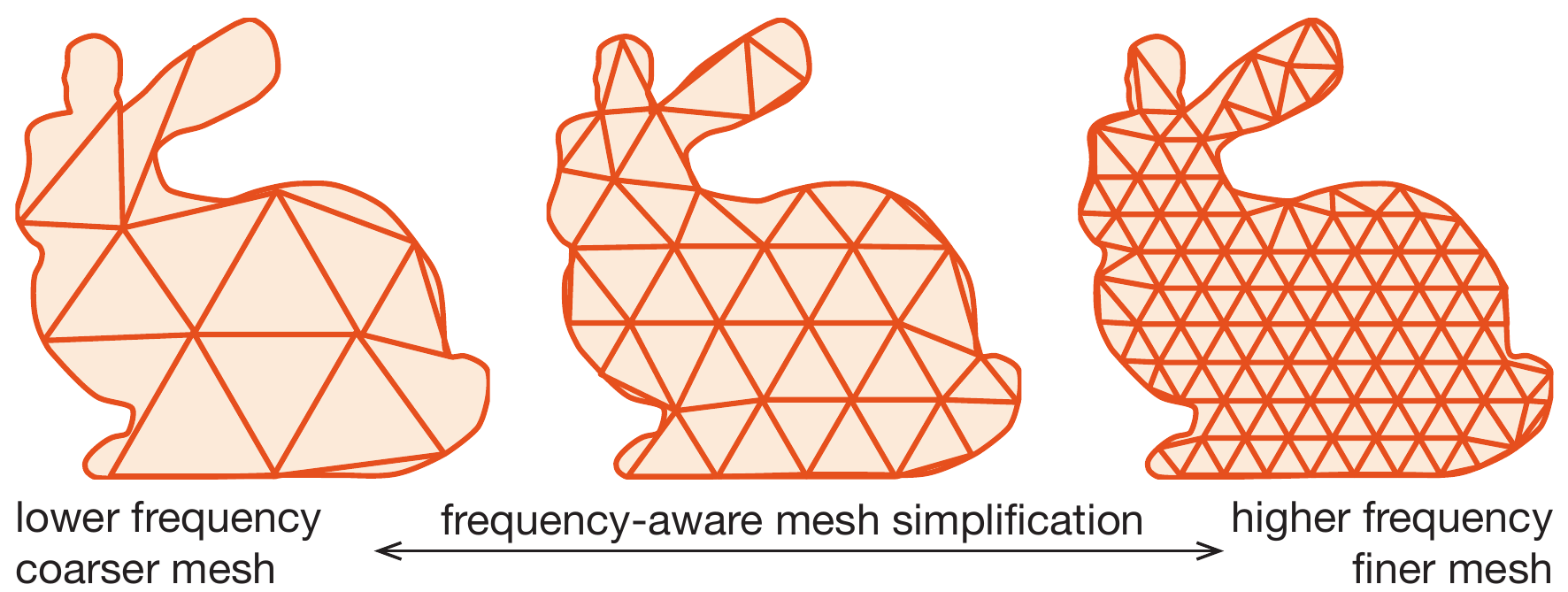}
  \caption{
	The key idea is to simplify mesh differently at different frequencies.
	By carefully choosing the decimating parameters, 
	my method provides 10$\times$ speedup over naive computation while
	preserving the accuracy of the wave propagation. 
  \label{fig:meshsimp}}
  \vspace{-4mm}
\end{figure}

\paragraph{Computational Design for Fabrication}
To design custom-shaped geometries with more complex physical phenomena, 
offline computational optimization is usually used in the design process.
Digital Mechanical Metamaterials proposes to embody mechanical movements into 3D printed 
objects using a modular method~\cite{IonWKB17}. 
Each of the modular cells is specially designed such that they can pass 
digital information when connected together.
Acoustruments introduced a passive acoustic-based mechanisms for interactive controls
on smartphones~\cite{laput2015acoustruments}. Through carefully designed tube geometries
and materials, an expansive dataset of design primitives is generated for easy construction.
To reproduce physical haptic interaction during fabrication, HapticPrint explores and 
builds a library of various patterns to different types of compliance~\cite{TorresCKP15}.
While this type of research is a promising direction, I would like to investigate how
to better combine computational design with intuitive tools for the users.

\section{My Thesis Research}

My research lies mainly at the intersection of physics-based simulation and interactive tools
for computational designs. 
I have focused on developing efficient algorithms that lead to interactive tools for 
non-intuitive functional requirements.

\begin{figure}[t]
  \centering
  \includegraphics[width=0.8\columnwidth]{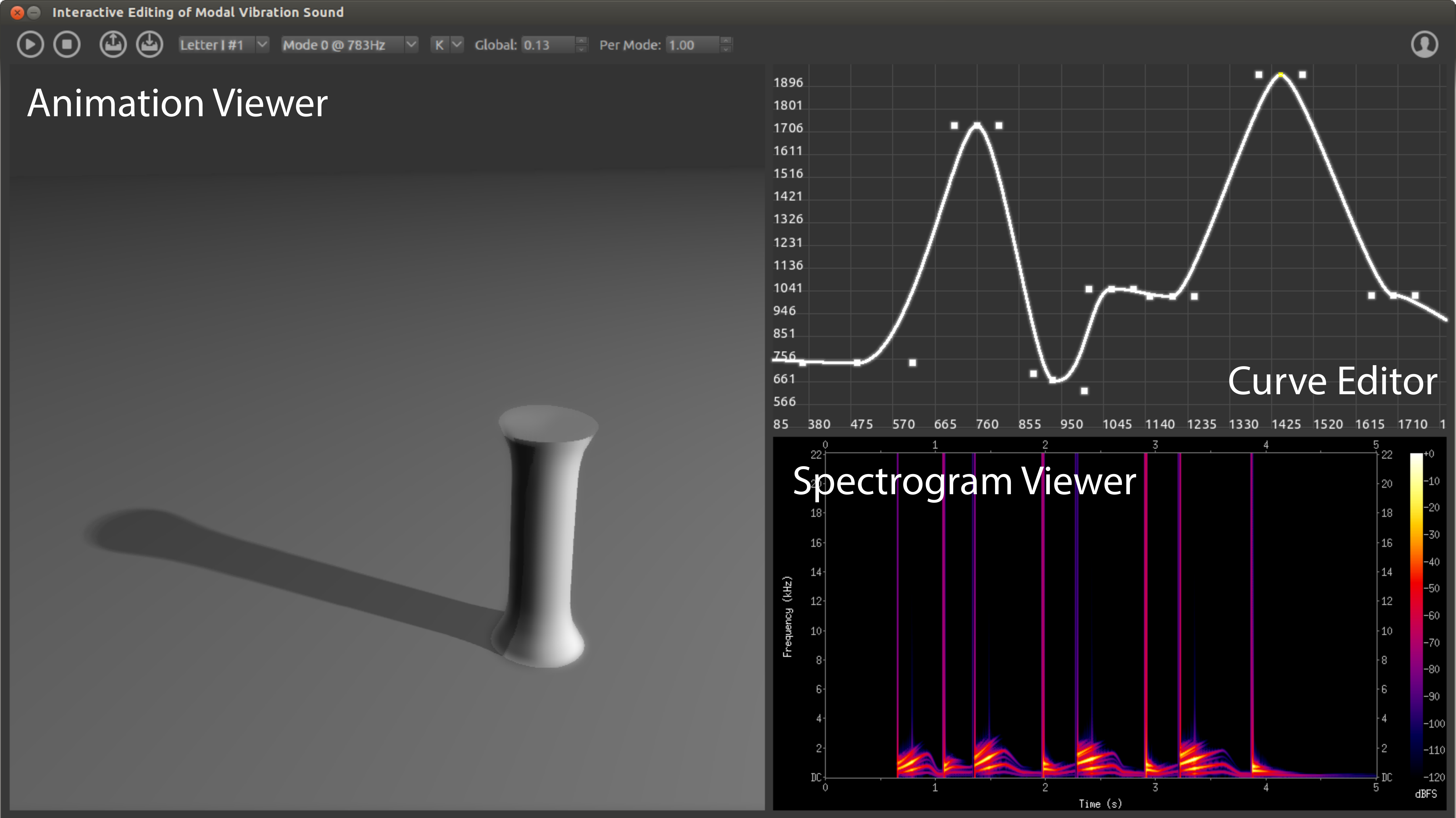}
  \caption{
	I designed and implemented an interactive sound editing interface.
    At runtime, an straightforward spline curve editor can be used 
    to edit the time-varying artistic effects.
    The sound propagation is precomputed using efficient physics-based 
    simulation and evaluated interactively at runtime.
  \label{fig:interface}}
  \vspace{-4mm}
\end{figure}

\subsection{Interactive Material-based Sound Editing}

Accurate audio in virtual reality is crucial for a fully immersive experience.
To edit virtual sounds from physics-based simulation, 
current sound models compute the modal vibrations and 
solve for the wave propagation at these vibration frequencies.
Figure~\ref{fig:teaser}-(a) shows some representative materials.
During the interaction, whenever the user tweaks material parameters,
the modal vibration frequencies changes completely.
At these new modal frequencies, expensive recomputation of sound propagation is required.
In my research, I developed a new system to speed up the computation,
enabling interactive and continuous editing as well as 
the exploration of material parameters~\cite{Li:2015:transfer}.

One of our key contributions is an efficient precomputation method for sound 
pressure fields.
Since precomputation is needed over a frequency range, the main bottleneck 
becomes the numerical solves of the wave equation.
It is known that the complexity of these depend on the number of surface elements $N$.
The smaller $N$ is, the faster computation can be.
The element size is bounded by the wavelengths at different frequencies.
Intuitively, the idea is to use coarser mesh while preserving the accuracy of the solves
(Figure~\ref{fig:meshsimp}).

The uniqueness of our work lies in the interactive material editing interface where
the users can freely explore at runtime, as shown in Figure~\ref{fig:interface}.
Our tool allows for interactive preview of the synchronized simulated sound that
reflects the user editing. 
The fast iteration enables users to explore artistic sound effects interactively
which would take minutes using naive implementation.

\subsection{Acoustic Voxels: Efficient Computational Fabrication}

Acoustic filters have numerous important applications, whether to produce
a desired sound pitch or to attenuate undesired noise. 
The applications range from wind instruments to mufflers and hearing aids.
When sound waves pass through a cavity, the filtered
frequencies are largely affected by the \emph{shape} of the cavity.
However, for all but the simplest cavity shapes, the influence of the shape on
the filtered frequency bands is complicated and unintuitive.

\begin{figure}[t]
  \centering
  \includegraphics[width=1.0\columnwidth]{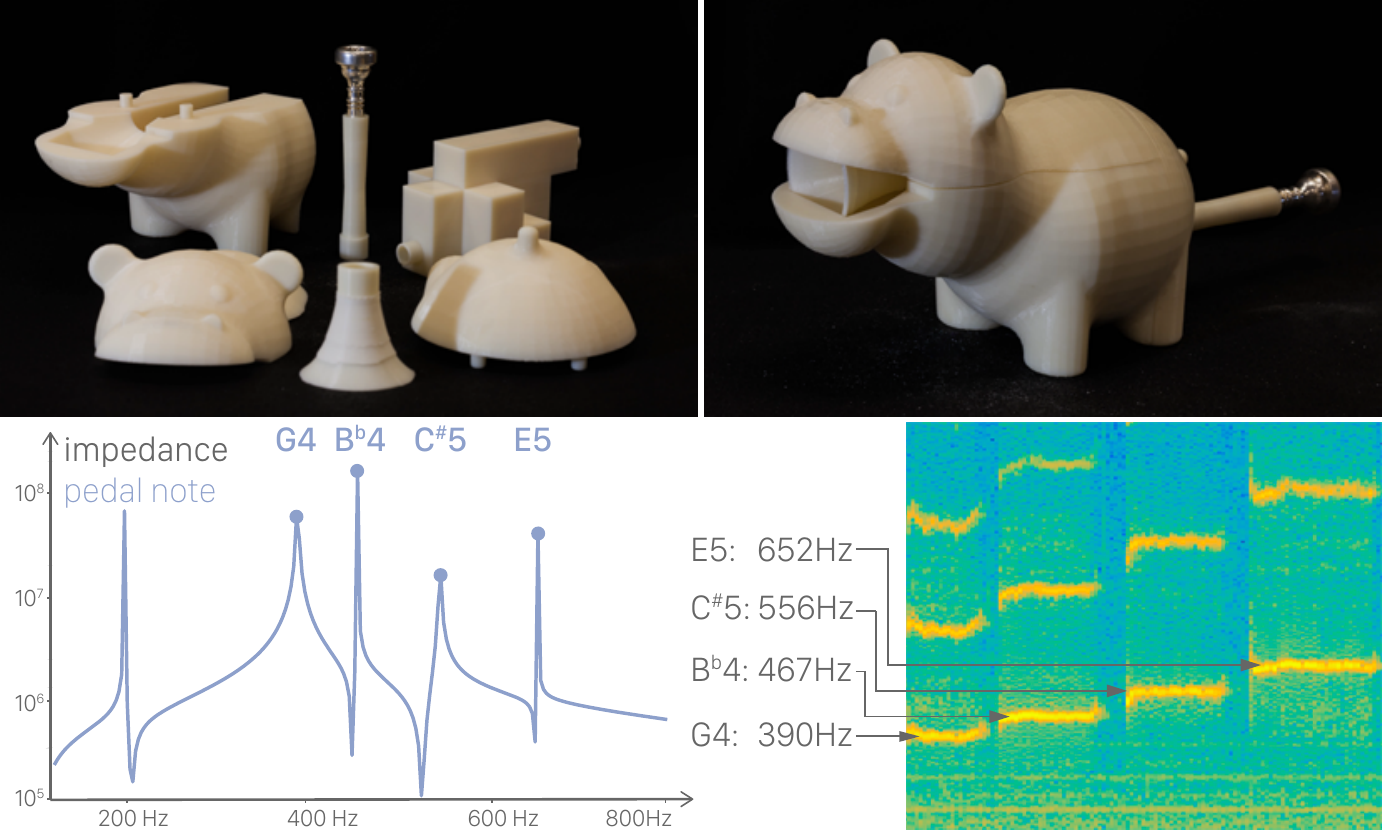}
  \caption{
  Acoustic Voxels help prototype  wind instruments in unconventional shapes by
  efficiently optimizing internal shapes, as shown in the top row.
  Provided with the desired pitches, our tool can generate a 3D mesh ready for printing.
  The recorded spectrogram at the bottom right shows the efficacy of our design tool.
   \label{fig:hippo}}
  \vspace{-4mm}
\end{figure}

I developed Acoustic Voxels, a computational tool that builds complex cavity shapes from
basic shape primitives.
The assembled cavity will produce the desired acoustic filtering effects. 
I proposed a modular scheme which not only simplifies the precomputation process on the 
primitive shapes but also drastically speeds up the design process.
Since typical numerical simulations scale nonlinearly with the number of elements, 
to predict the filtering behavior of a complex shape, 
it might take hours to simulate.
I demonstrated that the solving time in the proposed reduced design space can be 
reduced to seconds.

\begin{figure}[t]
  \centering
  \includegraphics[width=1.0\columnwidth]{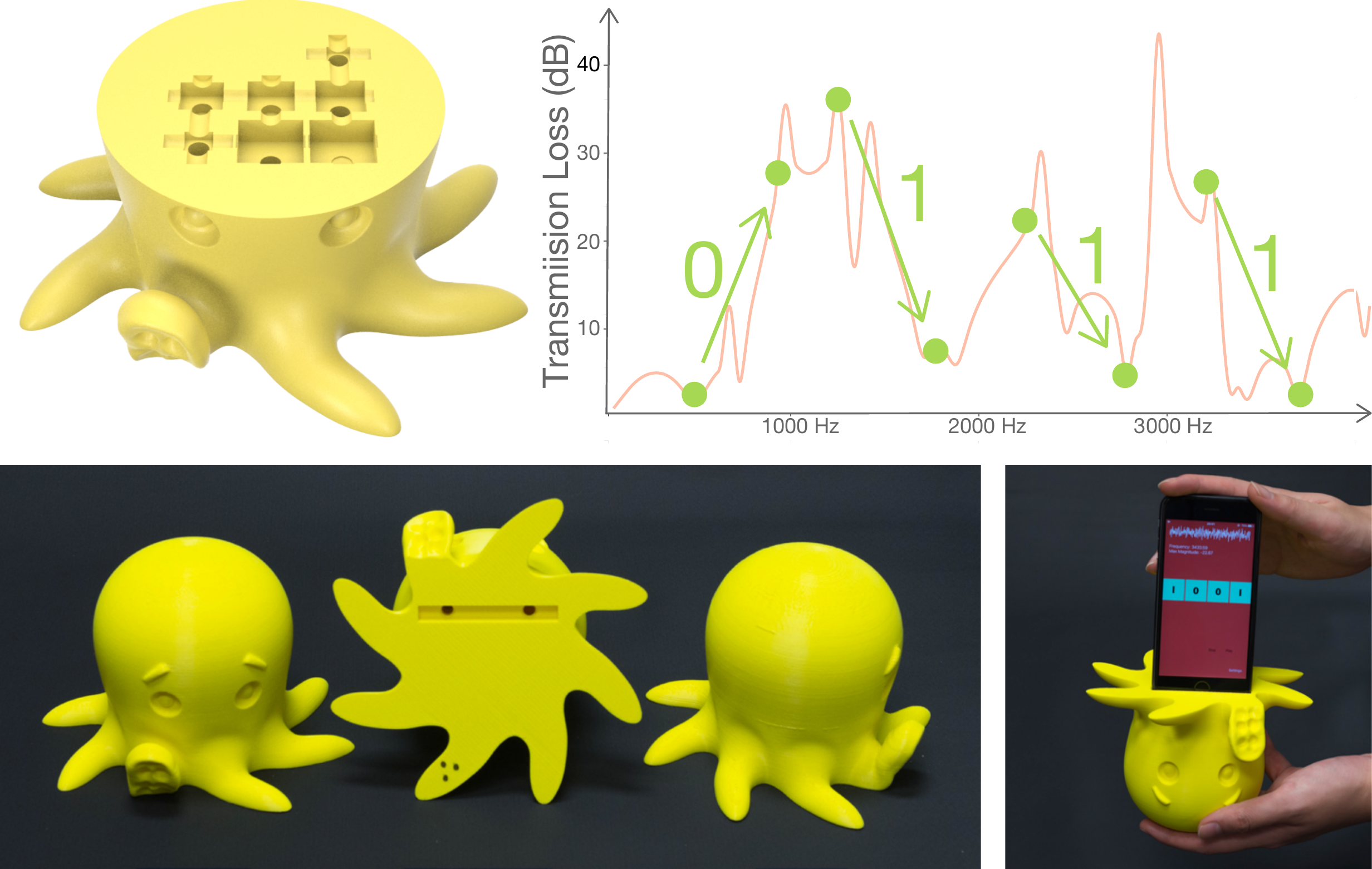}
  \caption{
  Example of acoustic encoding. 
  Acoustic Voxels optimizes and embeds voxels in the octopus.
  When interacting with an iPhone,  the iPhone app detects the encoded 4-bit binary data.
  \label{fig:oct}}
  \vspace{-4mm}
\end{figure}

I also presented a number of applications including wind  instrument prototyping, 
acoustic tagging, and acoustic encoding~\cite{Li:2016:acoustic_voxels}.
Figure~\ref{fig:hippo} illustrates a prototype wind instrument in a hippo shape,
enabled by our efficient and accurate simulation tool. 
Figure~\ref{fig:teaser}-b shows an example where we tagged piggy with desired acoustic signature which can be detected via tapping on the nose.
Taking tagging one step further, Figure~\ref{fig:oct} demonstrates the potential
to encode data in the acoustic filtering process.
Acoustic Voxels makes all these designs possible by 
utilizing efficient physics-based simulation, freeing users from dealing 
with non-intuitive physical requirements.

\subsection{AirCode: Unobtrusive Tagging for 3D Printing}

\begin{figure}[t]
  \vspace{-1mm}
  \centering
  \includegraphics[width=0.92\columnwidth]{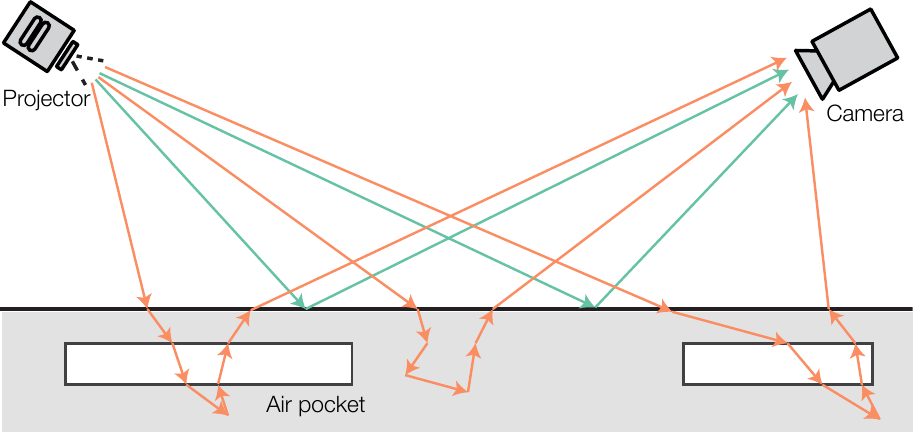}
  \caption{
  Key idea: Most plastic 3D printing materials
   exhibit strong subsurface scattering.
  Light rays (green) that are reflected by the surface represent the direct component;
  rays (orange) that enter the surface and are scattered within before leaving the surface result in the global component. 
  A structured change in the material that lies beneath the surface only affects the global component of a captured image.
  \label{fig:illu}}
   \vspace{-6mm}
\end{figure}

Motivated by the acoustic tagging application, I am interested in unobtrusive ways to
embed tags since acoustic filters require one inlet and one outlet (see the bottom of the octopus).
Taking advantages of subsurface scattering, 
I propose AirCode, an unobtrusive tagging tool for 3D printed objects.
As illustrated in Figure~\ref{fig:illu}, 
the key idea is simply placing thin air pockets under the surface of 3D printed objects. 
Air changes how light is scattered after penetrating the material surface. 
Most plastic 3D printing materials, even those considered opaque, scatter light.
The amount of light penetrating and scattered is often weak and 
most of the light is directly reflected at the surface. 
Consequently, the effects of air pockets on appearance can be made imperceptible to our eyes.
The scattered light can be separated out with a computational imaging method.

AirCode helps determine the shapes and positions of subsurface air pockets to 
encode useful information at a user-specified smooth region.
Intuitively, if the air pockets are very close to the surface, they are no longer imperceptible
to human vision. 
On the other hand, if the air pockets are put too deep to be discernible with an imaging system, 
it is impossible to extract and decode the embedded tags.
Combining the rendering algorithms and statistics from perception studies,
I designed a method to estimate a range of depths that satisfies our requirements.

\begin{figure}[b]
  \centering
   \vspace{-5mm}
  \includegraphics[width=0.8\columnwidth]{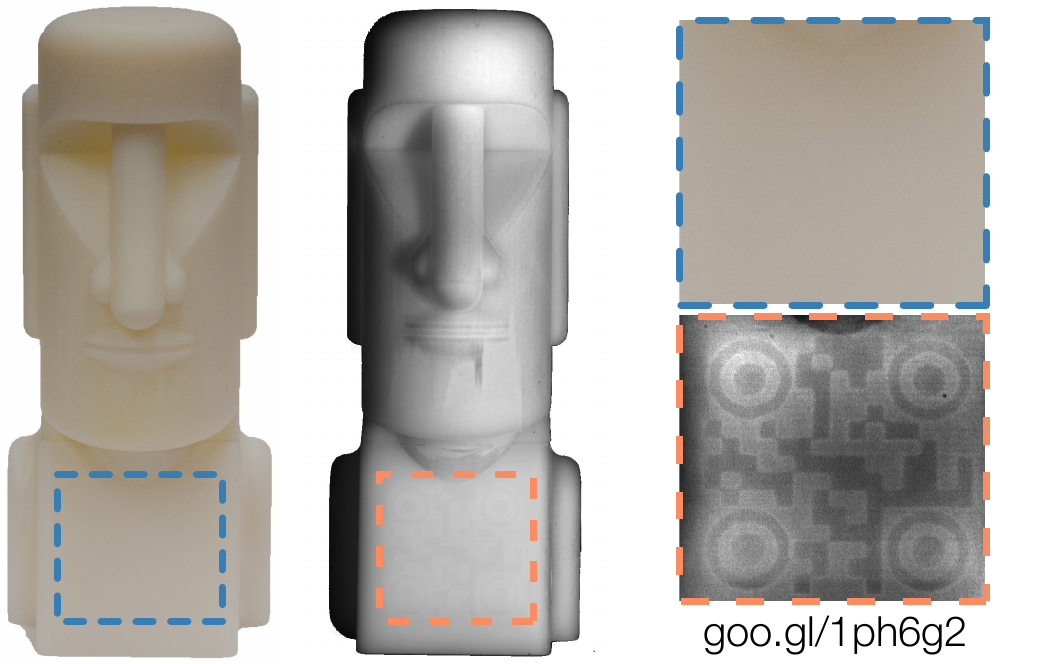}
  \caption{
  AirCode designs and embeds optimized air pockets beneath the surface.
  The fabricated model has the tag invisible under normal environmental lighting.
  The tag can be retrieved using our imaging system that separates out the 
   scattering effects. 
   In this example, we embedded a web link of the statue.   
   \label{fig:maoi}}
   \vspace{-4mm}
\end{figure}

One straightforward application is to embed  tags and metadata.
In Figure~\ref{fig:maoi}, an invisible tag is embedded in the statue and revealed
under computational imaging system, leading to a webpage with more details.
The embedded tag can not only provide additional metadata and digital identification 
but also  help estimate the pose of the object~\cite{aircode},
which becomes very helpful for robotic grasping.
If an object embeds AirCode tags, the camera system of a robotic manipulator
can recognize the object and retrieve its complete 3D model by reading the
tags. More remarkably, the located tags further allow the system to estimate the
object pose with respect to the camera.
With all the information, the robot gathers sufficient knowledge for a successful grasp.

\section{Future Work: \\Interactive Physics-based Design Tools}

To further integrate accurate simulation with interactive tools, 
there are a few directions that I want to explore.
First, I would like to work on high-fidelity sound simulations that run
at interactive rate.
Some of my past research has focused on the high-quality sounds. 
For example, we can simulate crumpling sounds when crushing a soda can or unwrapping a candy wrap
on powerful machines in the order of minutes or hours~\cite{Cirio:2016}.
Our work on efficient precomputation is a step towards more efficient simulations~\cite{Yang:2015:fastprecomp}.
I believe a general interactive pipeline that supports physics-based simulation
can bring the audio editing and design to the next level.

I am also interested in bringing rich statistics from recorded audios into simulations.
More specifically, I think a hybrid method between physics-based algorithms 
and data-driven statistics is a promising research area.
For a fully immersive virtual environment, while simulations can supply most of the
surrounding audios, there are certain subtle effects that are hard to simulate well.
For example, realistic room acoustic effects for indoor and natural ambient sounds
for outdoor scenes.
To this end, data-driven methods can fill in the gap by augmenting simulated results
with rich and real statistics from recordings.
I think combining these two themes of methods can further improve the quality
of VR audio.

In my past research fabrication projects, I mainly focused on how to embed metadata
or tags in a given 3D geometry through computational optimization. 
The use of physics-based simulation greatly eases the design process for the users.
In the future, I would like to explore more on the computational side of fabrication,
bringing more intuitive tools for ordinary users.
One idea is to use properties that are specific to certain printing processes. 
In AirCode project, I exploited the optical transparency in PolyJet printing material to embed tags.
However, for most FDM printers, the layer height is also an important parameter that has been
largely overlooked. 
I wish to build interactive tools that shows how  this parameter affects printing time 
and the strength of printed models. 
Furthermore, it is interesting to investigate how to embed information in varying layer heights.
I look forward to more interactive physics-based tools that can
enable more creative and functional designs.


%

 \section{Acknowledgments}
 
I would like to thank my advisor Changxi Zheng for his support and mentorship.
This research was funded in part by the NSF CAREER-1453101 and donations from Adobe and Intel. 
Dingzeyu Li was partially supported by a Shapeways EDU Grant and an Adobe Research Fellowship.

\balance{}

\bibliographystyle{SIGCHI-Reference-Format}
\bibliography{dli}

\end{document}